\documentclass[%
 onecolumn,
superscriptaddress,
 amsmath,amssymb,
 aps,
longbibliography
]{revtex4-2}

\usepackage{graphicx}% Include figure files
\usepackage{dcolumn}% Align table columns on decimal point
\usepackage{bm}% bold math
\usepackage{hyperref}% add hypertext capabilities
\usepackage[mathlines]{lineno}% Enable numbering of text and display math
\usepackage{soul}
\usepackage{float}

\renewcommand{\vec}[1]{\bm{\mathrm{#1}}}

\def \div{\nabla \cdot \mbox{}}
\def \grad{\nabla}
\def \lap{\nabla^2}

\def \x{\vec{x}}
\def \u{\vec{u}}
\def \s{\vec{s}}
\def \f{\vec{f}}
\def \F{\vec{F}}

\def \X{\vec{X}}
\def \U{\vec{U}}

\def \Ds{{\mathrm d}\s}

\def \Dx{{\mathrm d}\x}

\usepackage[usenames, dvipsnames]{color}
\begin{document}

\preprint{APS/123-QED}

\title{Pairwise and collective behavior between model swimmers at intermediate Reynolds numbers}% Force line breaks with \\

\author{Thomas Dombrowski}
%\thanks{Contributed equally}
\affiliation{Department of Physics, The University of North Carolina at Chapel Hill, Chapel Hill, North Carolina 27599, USA}

\author{Hong Nguyen} 
\affiliation{Department of Applied Physical Sciences, The University of North Carolina at Chapel Hill, Chapel Hill, North Carolina 27599, USA}

\author{Daphne Klotsa} 
\email{dklotsa@email.unc.edu}
\affiliation{Department of Applied Physical Sciences, The University of North Carolina at Chapel Hill, Chapel Hill, North Carolina 27599, USA}

\date{\today}% It is always \today, today,

\begin{abstract}
We computationally studied the pair interactions and collective behavior of asymmetric, dumbbell swimmers over a range of intermediate Reynolds numbers and initial configurations. Depending on the initial positions and the Re, we found that two swimmers either repelled and swum away from one another or assembled one of four stable pairs: in-line and in-tandem, both parallel and anti-parallel. When in these stable pairs, swimmers were coordinated, swum together, and generated fluid flows as one.
We compared the stable pairs' speeds, swim direction and fluid flows to those of the single swimmer. The in-line stable pairs behaved much like the single swimmer transitioning from puller-like to pusher-like stroke-averaged flow fields. In contrast, for the in-tandem pairs we discovered differences in the swim direction transition, as well as the stroke-averaged fluid flow directions. Notably, the in-tandem V pair switched its swim direction at a higher $\text{Re}$ than the single swimmer while the in-tandem orbiting pair switched at a lower $\text{Re}$. We also studied a system of 122 swimmers and found the collective behavior transitioned from in-line network-like connections to small, transient in-tandem clusters as the Reynolds number increased, consistent with the in-line to in-tandem pairwise behavior. Details in the collective behavior involved the formation of triples and other many-body hydrodynamic interactions that were not captured by either pair or single swimmer behavior. Our findings demonstrate the richness and complexity of the collective behavior of intermediate-$\text{Re}$ swimmers.

\begin{description}
\item[PACS numbers]
May be entered using the \verb+\pacs{#1}+ command.
\end{description}
\end{abstract}

\pacs{Valid PACS appear here}% PACS, the Physics and Astronomy
\maketitle

\section{\label{sec:Intro}Introduction}

In nature a large number of organisms at various scales use the surrounding medium fluid to perform vital functions, such as to move, feed and mate. Organisms also benefit also from being in pairs or groups as they increase swimming efficiency and speed,  feeding rates, reproductive processes, and social interaction~\cite{xu2019self,houghton2018vertically,chisholm2018partial,wilhelmus2014observations,kelley2013emergent}. 

Locomotion in the microscopic world is largely limited by the so-called scallop theorem, which states that a swimmer with a reciprocal, time-reversible swim stroke cannot produce a net motion in Stokes flow~\cite{Purcell1977}.
Thus, real microscopic organisms break time reversibility with non-reciprocal swim strokes. 
Moreover, it is rare to find a swimmer isolated from other organisms, which has led to numerous studies (both theoretical and experimental) on the hydrodynamic interactions between Stokesian swimmers~\cite{sokolov2015individual,elfring2011passive,yang2008cooperation,ishikawa2006hydrodynamic,kanevsky2010modeling,lippera2020bouncing}. For example, sperm cells and B. Subtilis synchronize their flagella, and in doing so are hydrodynamically attracted to one another~\cite{sokolov2015individual,elfring2011passive,yang2008cooperation}. Many microscopic swimmers can be modeled as squirmers and then be classified as either a ``puller'' or a ``pusher''; the former has  a force dipole, which induces a flow field that pulls in parallel to its swimming axis and pushes out in the perpendicular, while the latter does the opposite. Pullers have been shown to attract along their swimming axis and repel in the perpendicular and pushers have been shown to repel along the swimming axis and attract along the perpendicular~\cite{lauga2009hydrodynamics,ishikawa2006hydrodynamic,kanevsky2010modeling}. Collision dynamics for a model of chemically active droplets were shown to be sensitive to the relative sizes of the droplets and thus to affect the subsequent dynamics~\cite{lippera2020bouncing,lippera2021alignment}. 

The scallop theorem does not have to hold when there is more than one swimmer (and thus time-reversibility can be broken by the presence of the additional swimmer)~\cite{koiller1996problems}. For instance, when two reciprocal, identical dumbbell swimmers oscillate with a phase difference, they could interact to produce a net motion like a single non-reciprocal swimmer. The dumbbell is a common model for a simple reciprocal swimmer composed of two spheres that oscillate with respect to each other. 
Studies on the hydrodynamic interactions between a pair of dumbbell swimmers in Stokes flow have shown how swimmers align and swim together, depending on their initial configuration~\cite{alexander2008dumb,lauga2008no,Putz2010}.

The pair interactions between two asymmetric dumbbells have been analytically solved for the one-dimensional case of reflection invariant pairs (collective swimming at a constant velocity) and translation invariant pairs (attractive or repulsive dependent on the leading swimmer orientation)~\cite{lauga2008no}. A more extensive study developed stroke-averaged equations of motion for the effective hydrodynamic interactions between asymmetric dumbbell pairs in Stokes flow~\cite{Putz2010}.

Previous studies of interacting swimmers have been mostly limited to the Stokes regime, where inertial forces are negligible. However, as the Reynolds number increases, the inertial contribution to the system's dynamics becomes important. Indeed, inertial effects on a model mesoscale swimmer can induce a switch in the swimming direction~\cite{dombrowski2019transition,dombrowski2020kinematics} and impact predator/prey dynamics~\cite{gemmell2013compensatory}. Recently, there has been a growing interest in understanding how inertial effects impact interactions between squirmers~\cite{gotze2010mesoscale,li_arezoodaki_2016hydrodynamic}. Squirmers aligned initially in parallel will attract if they are pushers and repel if they are pullers~\cite{gotze2010mesoscale}, and when on a collision course, inertial effects change the contact time and dynamics for pushers and prompt hydrodynamic attraction for pullers~\cite{li_arezoodaki_2016hydrodynamic}. Less is known about other inertial swimmers, models beyond squirmers or real ones. Additionally, the role of inertia on the collective behavior of swimmers beyond a pair is largely unexplored, with the exception of Chatterjee \textit{et al.} who identified stable, unstable and turbulent states for active suspensions of weakly inertial pushers~\cite{chatterjee2019fluid}. 

In this paper, we computationally studied the pairwise and many-body hydrodynamic interactions for model reciprocal, asymmetric dumbbell swimmers over a range of intermediate Reynolds numbers. Varying the initial positions and orientations for two swimmers, we found regions where they repel and swim away from one another and regions where they interact to form stable pair configurations. From thousands of initial conditions, only four stable pairs were identified, which can be grouped as: in-line and in-tandem, parallel and anti-parallel. Parallel in-tandem pairs form a V-shape and antiparallel form a dynamic orbit.  In stable pairs, swimmers were coordinated, swum together, and generated fluid flows as one. We studied the pairs' combined fluid flows and velocities as a function of the Reynolds number. We found that in-line pairs act similar to a single swimmer and are more frequent at low Re, while the in-tandem pairs have more complex dynamics and become more frequent as Re increases. In-tandem pairs show a transition from small-sphere-leading to large-sphere-leading coordinated swimming at different Reynolds numbers to each other and different compared to the single swimmer~\cite{dombrowski2019transition}. 
Finally, we simulated 122 swimmers and found a transition from an assembly of in-line network-like connections to in-tandem transient clusters as the Reynolds number increased. Pairwise interactions were used to partly explain the collective behavior; however, limitations were discovered as many-body interactions such as triples were also identified.

The structure of the paper is as follows. In section II, we describe the model, computational method, and simulation details. In section III we present our findings on pair behavior: A) the stable pairs formed by two asymmetric dumbbells, B) the comparison between swimming as an individual and as a stable pair, and C) the stroke-averaged fluid flows. In section IV we discuss the evolution of the collective behavior and end with conclusions in section V.
%We end with the evolution of the collective behavior in section IV and conclusions in section V.

\begin{figure*}
\includegraphics[width=0.95\textwidth]{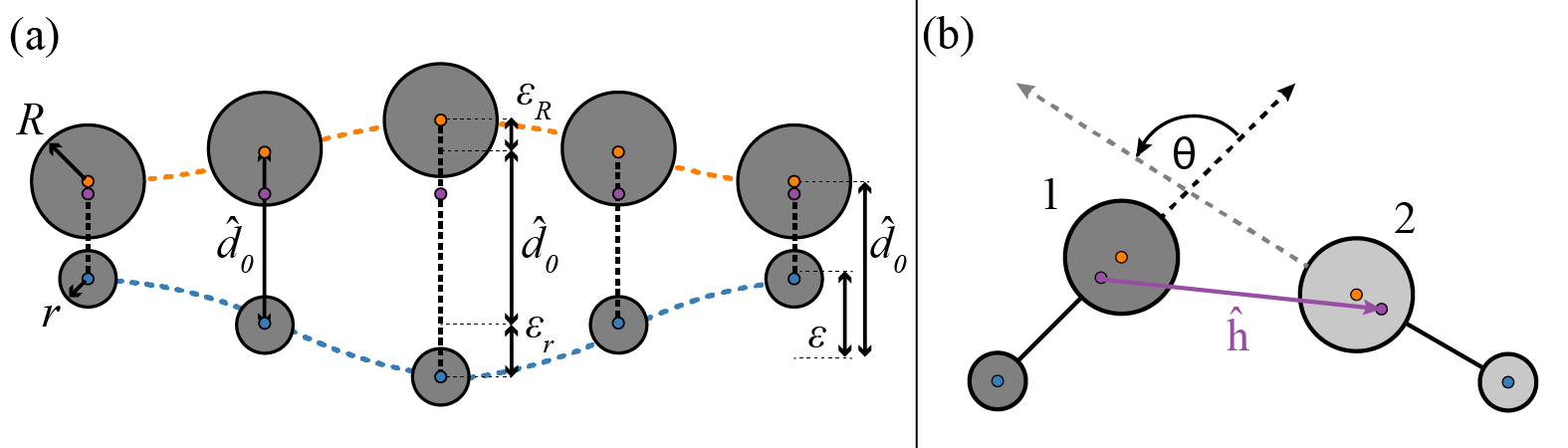}
\centering
\caption{(a) Swim stroke of the asymmetric dumbbell with $\alpha = 0.5$. The dumbbell's large sphere (orange) oscillates in antiphase with the small sphere (blue) and the distance between them is prescribed to be of a simple harmonic oscillator with equilibrium length $\hat{d}_0$ and amplitude $\varepsilon$. (b) A pair of asymmetric dumbbells separated by a distance $\hat{h}$ and oriented with angle $\theta$. $\theta$ is formed by the counter-clockwise rotation from the swimming axis of swimmer 1 to that of swimmer 2.}
\label{SinglePair}
\end{figure*}

\section{Model and Methods}

We consider a two-dimensional system of identical, asymmetric dumbbell swimmers. Each dumbbell is composed of two unequal sized spheres of radii $R$ and $r$ with an aspect ratio of $\alpha = r/R = 0.5$, see Fig~\ref{SinglePair}a. The dumbbell's swimming axis is along the line connecting the centers of the two spheres. The following conditions ensure that the dumbbell swimmer is force- and torque-free. The spheres are neutrally buoyant with respect to the surrounding fluid. The distance between the two spheres $\hat{d}(t)$ is prescribed to be of a simple harmonic oscillator such that $\hat{d}(t) = \hat{d}_0 + \varepsilon\text{sin}(2\pi \tau)$, where $\hat{d}_0 = d_0/R = 2.5$ is the equilibrium distance between sphere centers, $\varepsilon = 0.8$ is the amplitude of oscillation, and $\tau = ft$ is the dimensionless time parameterized by $f$ the frequency of oscillation. The amplitudes of each sphere are determined by the conservation of momentum $\varepsilon_r = \varepsilon R^2/(r^2+R^2)$ and $\varepsilon_R = \varepsilon r^2/(r^2+R^2)$, where subscripts $R$ and $r$ represent quantities specific to the large sphere and small sphere, respectively. Because $R=2r$ we find that $\varepsilon_r = 4\varepsilon_R$.

The asymmetric dumbbell swimmers were immersed in a viscous, incompressible Newtonian fluid that occupied a finite cell with periodic boundary conditions. The fully-coupled fluid-structure interaction system was resolved using 
the Immersed Boundary (IB) method~\cite{griffith2007adaptive,griffith2012immersed}. The IB scheme was implemented in the open-source IBAMR software, which is an immersed boundary numerical method with adaptive mesh refinement~\cite{griffith2007adaptive}. 
The IB method for fluid-structure interaction uses an Eulerian formulation of the momentum equation and incompressibility constraint for the coupled fluid-solid system along with a Lagrangian description of the motion of the immersed structures.  
Let $\x \in \Omega$ be fixed Eulerian physical coordinates, and let $\s \in U$ be fixed Lagrangian curvilinear coordinates attached to the structure, where $\Omega$ and $U$ are the physical regions occupied by the fluid-structure system and structures respectively.
In our notation, $\X(\s,t) \subset \Omega$ is the physical position of material point $\s$ at time $t$.
The momentum equation and incompressibility constraint are given by
\begin{eqnarray}
\rho \frac{D \u}{D t}(\x,t) &=& -\grad p(\x,t) + \mu \lap \u(\x,t) + \f(\x,t), \\\label{eqn_momentum}
\div \u(\x,t) &=& 0, \label{eqn_continuity}
\end{eqnarray}
in which $\u(\x,t)$ is the Eulerian velocity field, $p(\x,t)$ is the pressure field that imposes the incompressibility constraint, $\f(\x,t)$ is a body force that arises from the presence of the immersed structure, $\rho$ is the mass density, and $\mu$ is the viscosity.
Eulerian and Lagrangian variables are coupled via integral transforms with Dirac delta function kernels:
\begin{eqnarray}
\f(\x,t)   &=&  \int_{U} \F(\s,t) \, \delta(\x - \X(\s,t)) \, \Ds, \label{eqn_F_f} \\
\U(\s,t)   &=&  \int_{\Omega} \u(\x,t) \, \delta(\x - \X(\s,t)) \, \Dx \,. \label{eqn_u_U}
\end{eqnarray}
Eq.~(\ref{eqn_F_f}) converts the Lagrangian force density $\F(\s,t)$ to an equivalent Eulerian force density $\f(\x,t)$, and Eq.~(\ref{eqn_u_U}) evaluates the local material velocity at each structural position.

In our computer model, each sphere of the dumbbell was discretized using a collection of Lagrangian marker points that were generated using an in-house Python code, and the singular delta function kernels were replaced by a four-point regularized kernel function~\cite{griffith2007adaptive}. To maintain each sphere's rigidity, each marker point was connected with their nearest neighbors via intra-sphere stiff springs. 
The spheres were also connected by a set of inter-sphere springs which controlled their swim stroke oscillation and prevented individual sphere rotation. The force $\F(\s,t)$ applied on marker point $\s$ at time $t$ was solely due to the linear expansion/compression of the springs. The spring force for marker points $\s_1$ and $\s_2$ connected by the spring $\ell$ was given by:
\begin{eqnarray}
\F^\ell(\s_1,\s_2,t) &=& -K_s(|\X(\s_1,t) - \X(\s_2,t)| - R_{\ell}) \label{eqn_f_spring}
\end{eqnarray}
where $K_s$ was the spring stiffness coefficient and $R_{\ell}$ was the resting length of spring $\ell$. We note that $\F^\ell(\s_1,\s_2,t) = -\F^\ell(\s_2,\s_1,t)$. The resting length of the inter-sphere springs updates to the prescribed distance between $\s_1$ and $\s_2$ at every time step, while the resting length for the intra-sphere springs is kept fixed. For the inter-sphere springs, $K_s = 1.0\times10^4$N/m, and for the intra-sphere springs, $K_s = 5.0\times10^4$N/m. These stiffness coefficients have been chosen small enough for numerical stability and also large enough to ensure that the individual sphere's deformation is negligible, and that the distance between spheres is approximately kept at the prescribed distance at all time.  More details on the spatial discretization and the time stepping algorithm for the IB method can be found in references ~\cite{griffith2007adaptive,griffith2012immersed}.

In IBAMR, an adaptive fluid grid is implemented to improve the efficiency of the simulation. There were four refinement levels (N=16, 64, 256, and 1024) and the dumbbell meshes were evaluated at the highest grid refinement of N=1024 and grid spacing of $\hat{d\text{X}} = d\text{X}/R = L/NR = 0.049$, where $L$ is the size of the simulation box. The spacing between marker points is set by the standard IB method to be $\hat{d\text{S}} = 0.5\hat{d\text{X}}$ to avoid fluid leak into the spheres. 
The simulation box $L$ was large enough to prevent finite size effects, $L=12.5(d_0+R+r)=50R$.
Close contact between immersed structures are automatically handled by IBAMR with an enhanced version of the kernel function~\cite{Griffith2009Immersed}. Therefore, no special treatment is needed for the collision between dumbbells within the IB scheme currently employed.

The pair system was composed of two identical dumbbells parameterized by the separation distance at time $t$, $\hat{h}(t)= h(t)/R$ between the dumbbells' centers of mass, and the counter-clockwise angle formed between their swimming axes $\theta$, see Fig.~\ref{SinglePair}b. The initial conditions are stated as $\hat{h}(t=0)= \hat{H}= \langle \hat{H}_x,\hat{H}_y \rangle$  We used the frequency Reynolds number of the small sphere $\text{Re} = \alpha\varepsilon_rM^2$, with viscosity $M^2 = R^2/\delta^2$ and boundary layer thickness $\delta = \sqrt{\nu/\omega}$. The reason for the choice of Re is that it has been found to control the transition of swimming dynamics in the single dumbbell's system~\cite{dombrowski2019transition,dombrowski2020kinematics}. Note that there are many Reynolds numbers (dimensionless ratios) that can be defined in this system because of the many length scales -- for a discussion see Dombrowski \textit{et al.}~\cite{dombrowski2020kinematics}. We monitored the swimming pair until the swimmers either diverged ($\hat{h} > 10$) or reached a steady state (separation distance $\hat{h}$ and angle $\theta$ changed by less than one percent over consecutive swim strokes). We monitored the pair systems until steady state had been reached, and the simulation duration varied from at least twenty to over four hundred oscillations.

\begin{figure*}
\includegraphics[width=1\textwidth]{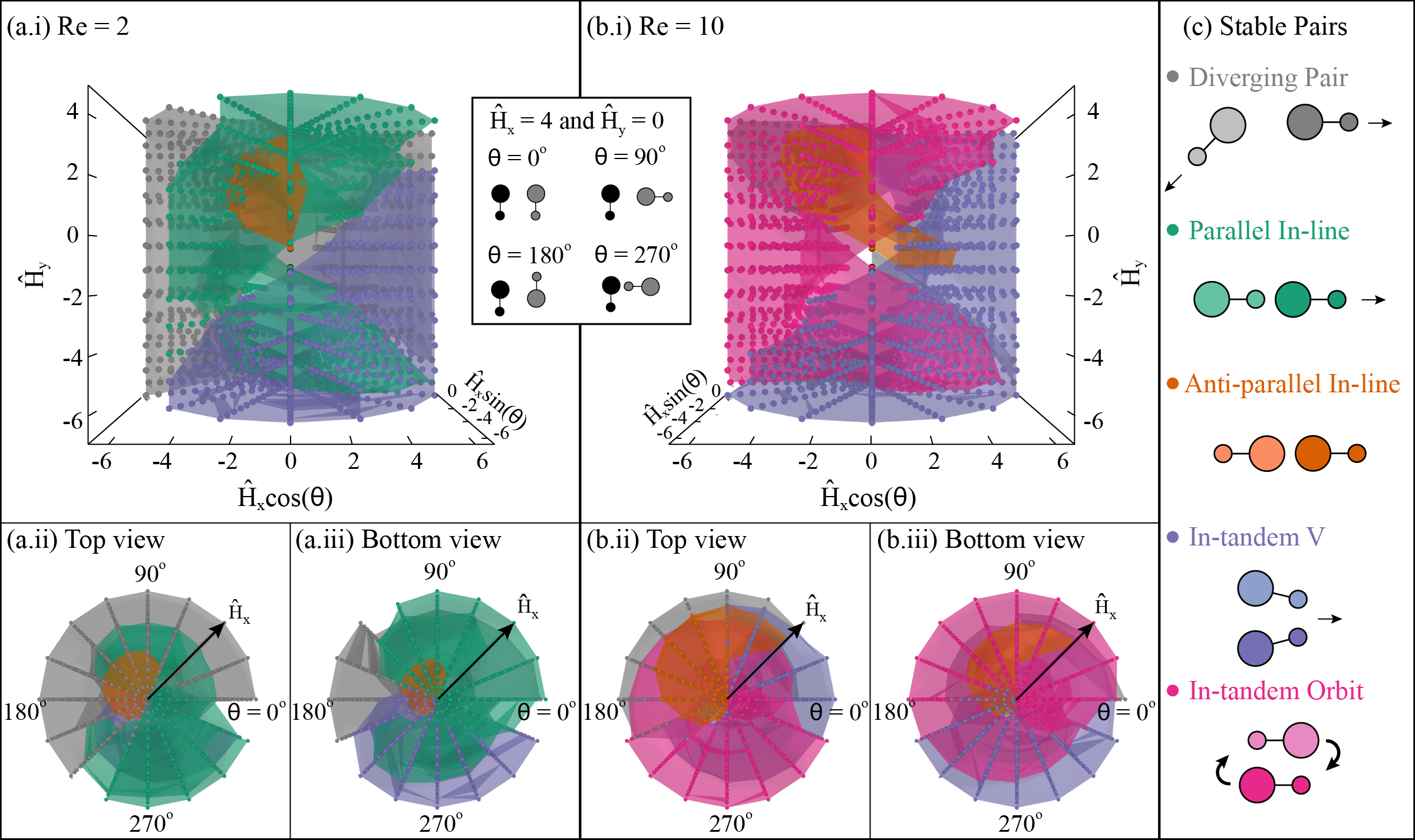}
\centering
\caption{Identified regions of diverging or stable pairs after steady state had been reached for (a) $\text{Re} = 2$ and (b) $\text{Re} = 10$ based on initial configurations $\hat{H}_x$,$\hat{H}_y$, and $\theta$. Initial configurations are determined based on the position and orientation of the grey swimmer with respect to the black swimmer. (ii) Regions of pair behavior when the grey swimmer is above the black swimmer and (iii) when the grey swimmer is below the black swimmer. (c) Schematic of pair behaviors: diverging (grey), parallel in-line (green), anti-parallel in-line (orange), in-tandem V (blue), and in-tandem orbit (pink).}
\label{InitConfig}
\end{figure*}

When extending to multiple swimmers, the IB method becomes computationally expensive due to the large number of marker points, stiff springs, and increasingly smaller time steps. Therefore, in the supplemental information (SI), we describe an alternative computational scheme with which we simulated a system of 122 swimmers.

\section{Pair Results}
\subsection{Pair stable states}

We first investigated how two swimmers interacted starting from different initial configurations, for three representative intermediate Reynolds numbers, $\text{Re} = 2$, $7$, and $10$. Specifically, we performed a large sweep (more than 5,600 simulations) over position and rotation space, varying the initial separation distance, $0.25 \leq \hat{H_x} \leq 6.25$ and $-6.5 \leq \hat{H}_y \leq 4.5$, over angles $\theta$ in the range $0-360$ in increments of $360/16$. 

We found that for all initial configurations, and Reynolds numbers studied, the swimmers either swam away from one another or converged to one of only four stable pair states, that are the combinations of in-line and in-tandem, parallel and antiparallel arrangements, as shown schematically in Fig.~\ref{InitConfig}(c). Once the swimmers arranged into their preferred stable pair, they then swum together in a coordinated motion as one. We plotted three-dimensional ``phase diagrams'' showing the resulting stable pairs as a function of the initial relative positions and angles between the swimmers. It is interesting to compare these diagrams for $\text{Re} = 2$ and $\text{Re} = 10$, see Fig.~\ref{InitConfig} (a),(b). For $\text{Re} = 2$, we see a large portion of the initial-conditions phase space assembles parallel in-line pairs or leads to diverging pairs, and there are also smaller regions of antiparallel in-line and in-tandem V pairs, see Fig.~\ref{InitConfig} (a). On the other hand, for $\text{Re} = 10$, we found that parallel in-line pairs have become in-tandem orbit pairs. Moreover, there is a larger region of in-tandem V pairs and almost no diverging pairs, see Fig.~\ref{InitConfig} (b). The $\text{Re} = 7$ phase diagram shows how the stable configurations gradually change as a function of Reynolds numbers between the $\text{Re}=2$ and $\text{Re}=10$, see SI (section III).  We note that there was a fourth state discovered which resembled an L shape. But this state was found to be meta-stable because all of the L shaped systems eventually transitioned into parallel in-line pairs. Next, we investigate each pair in more detail.

\subsection{Pair velocity as a function of Re}

\begin{figure}
\includegraphics[width=0.8\textwidth]{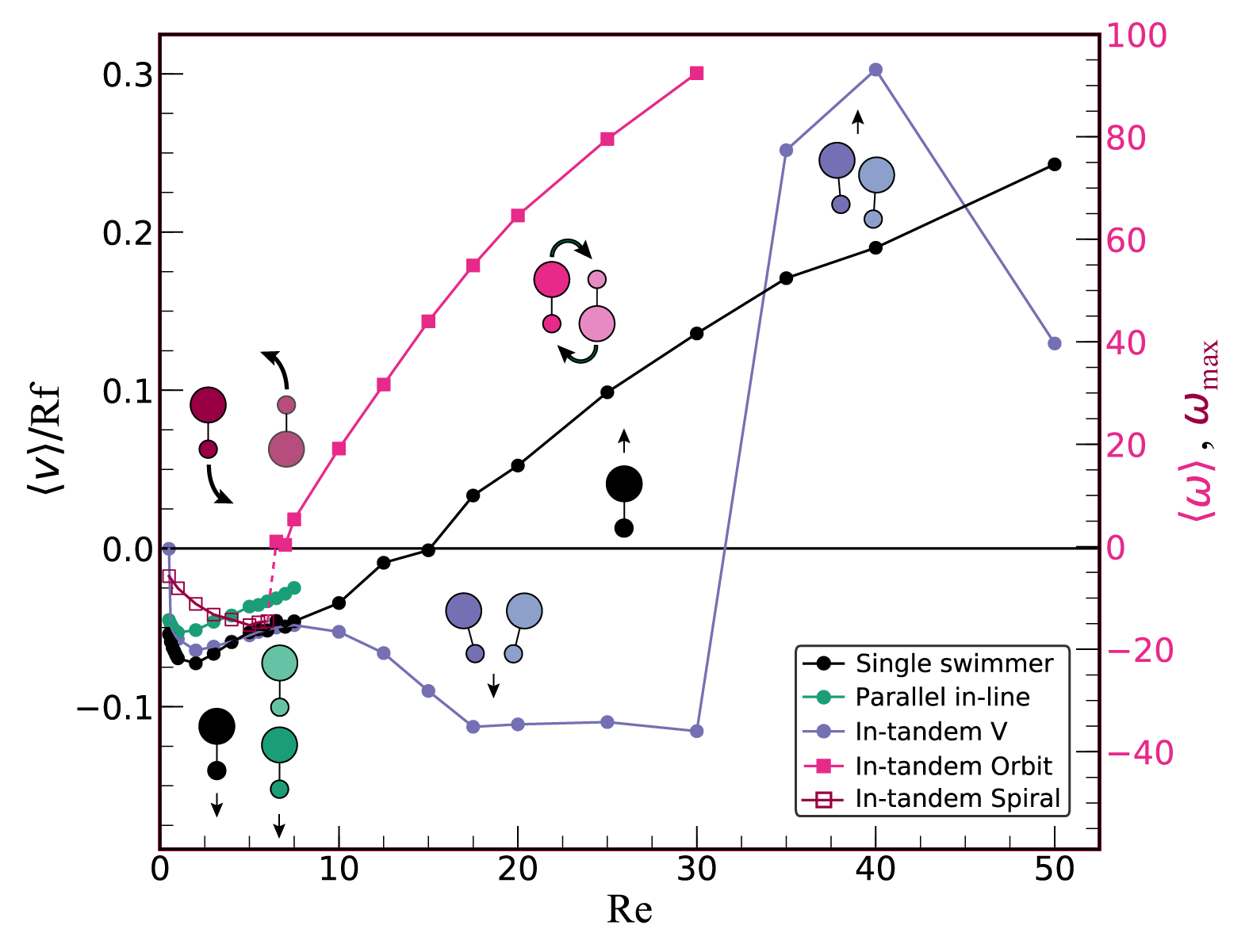}
\centering
\caption{The swimming velocity of an in-tandem V pair (blue), a parallel in-line pair (green), and a single swimmer (black), as well as the angular velocity of the in-tandem orbit pair (pink) versus the Reynolds number. The single swimmer switches its swimming direction from small-sphere-leading to large-sphere-leading at $\text{Re} \approx 15$, the in-tandem V pair at $\text{Re} \approx 35$ and the in-tandem orbit at $\text{Re} \approx 6.5$}
\label{PairSingSwim}
\end{figure}

We monitored the pairs' steady-state velocity, as well as their separation distance $H$ across $\text{Re}$ to understand how different pair dynamics compares to those of a single swimmer and thus provide insight on possible benefits behind forming each stable pair. We calculated the average velocity of the swimmers by averaging the velocity per oscillation over the total number of oscillations in steady state.  Note that, for the in-tandem orbit pair, we calculated the angular velocity. We excluded antiparallel in-line pairs from the comparison because they are stationary and have neither net translational nor angular velocity. A comparison of the velocities for the swimming and orbiting pairs and of a single swimmer are shown in Fig.~\ref{PairSingSwim}.
The pairs' separation distance as a function of Re is shown in the SI.

The parallel in-line pair swims small-sphere-leading, and the curve looks similar to the single swimmer in the Reynolds number range $0.5 \leq \text{Re} \leq 7.5$, but with a slightly lower speed than the single swimmer case. It seems reasonable to ascertain that the speed difference is due to the presence of the second swimmer in tow behind the leading swimmer. The swimmer in tow may be increasing the resistance, which could be attributed to a net attraction/pulling motion between the swimmers, an increase in fluid flow motion behind the leading swimmer, or the mass added to the system by the second swimmer. For $Re > 7.5$, the swimmers collide and push away from one another, so there is no stable pair for which to calculate an average translational velocity.

The in-tandem V pair swims small-sphere-leading for a larger Reynolds number range $0.5 < \text{Re} \leq 30$ compared to the single swimmer. While from $Re=0.5$ to $Re=7.5$ the V-pair and the single swimmer's velocity curves are almost on top of each other; for Re>7.5, the curves diverge. The V-pair actually speeds up till it stops moving in a straight line and picks up a rotational velocity and moves on an arc (which results in the plateau of the translational velocity around $Re\approx 17$, see Fig.~\ref{PairSingSwim}). Then at $Re\approx35$ we see a sharp transition from small-sphere-leading to large-sphere-leading. In other words, the Reynolds number at which the switch in swim direction occurs is much larger for the pair ($\text{Re}=35$) than for the single swimmer ($\text{Re}=15$), which we would not have predicted. 
Looking closely, the switch in swimming direction for the V-pair happens when the swim stroke actually changes a little bit, to include a stagger motion, see movie in the Supplemental Material. When $\text{Re} > 50$ swimmers eventually physically collide and no longer form a stable pair.

The in-tandem orbit pair is stable only in the large-sphere-leading regime, which occurs around $Re\approx6.5$ (a lower Re than for the single swimmer which is at $Re\approx 15$). For $Re<6.5$ the swimmers swim small-sphere-leading and move away from one another in a small-sphere-leading spiral. For $\text{Re} \ge 6.5$ the angular velocity of the stable orbit pair increases monotonically with increasing Re.

\subsection{Pair fluid flows and swimming}

\begin{figure*}
\includegraphics[height=0.85\textheight]{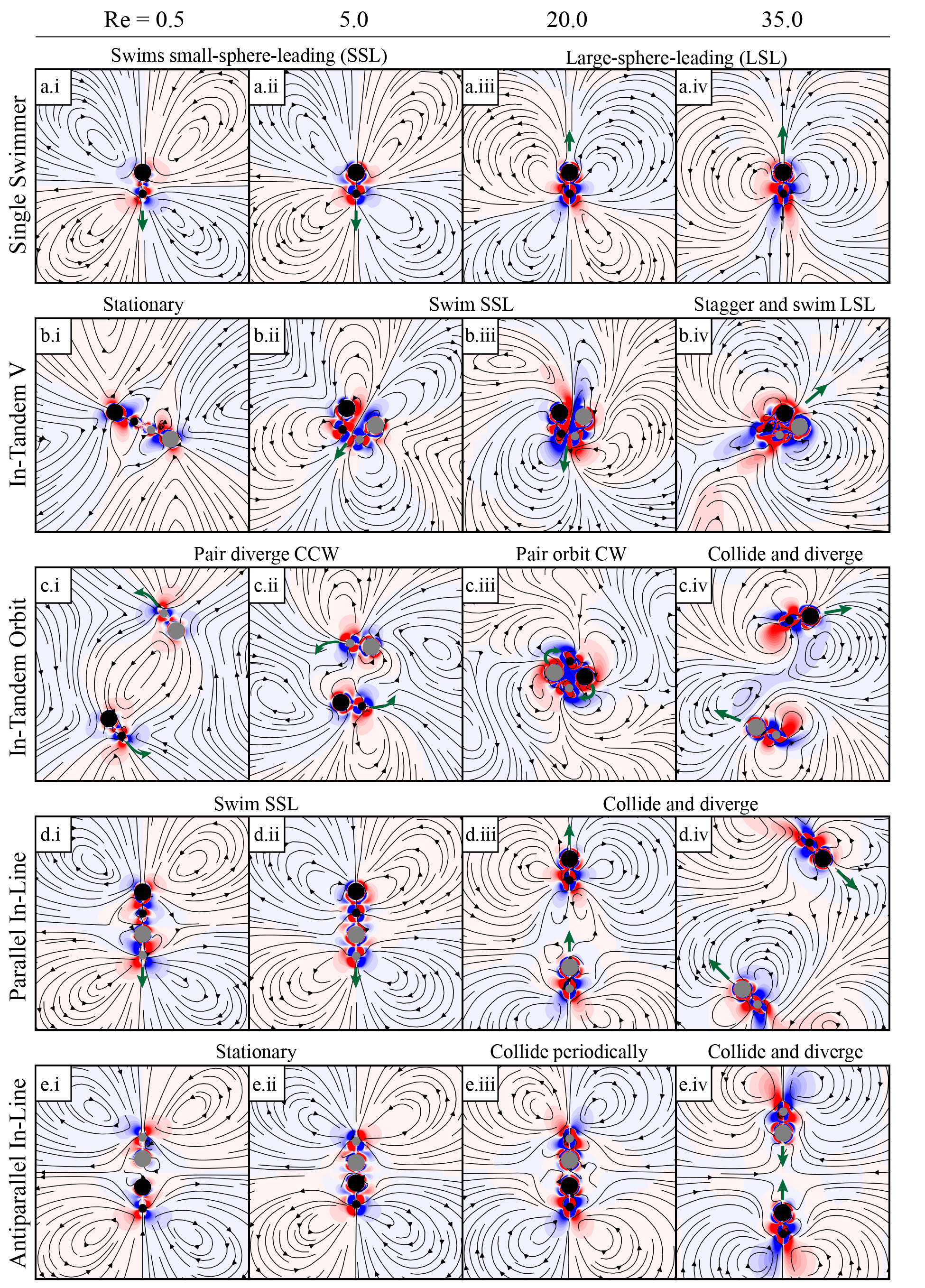}
\centering
\caption{Stroke-averaged fluid flow for (a) an individual dumbbell, (b) in-tandem V, (c) in-tandem orbit, (d) parallel in-line, and (e) anti-parallel in-line on the 100th swim stroke, $\tau = 100$.(a) Individual swimmer flow and swim direction transitions from puller-like SSL (i,ii) to pusher-like LSL (iii,iv). (b) In-tandem V forms a stationary pump pair, a V formation which swims SSL, and a staggering V formation which swims LSL. (c) In-tandem orbits diverge counter-clockwise when $\text{Re} < 6.5$ and form a stable orbit when $6.5 \leq \text{Re} < 30$. (d) Parallel in-line pairs form when $\text{Re} < 10$ and diverge otherwise. (e) Anti-parallel in-line pairs transition from stationary pumps to swimmers in a periodically occuring head-on collision.}
\label{FlowStream}
\end{figure*}

We then investigated the details of the stable pairs identified in the previous section and studied their fluid flow fields and how they changed as a function of the Reynolds number. Before we discuss the pair, we should explain what is known for the fluid flows of the single swimmer. In our previous work, we found that the dumbbell swimmer produces a time-averaged fluid flow (steady streaming) that changes as the Re increases. Specifically, for $\text{Re }< 15$ the flow field is puller-like, i.e. it pulls in along the swimming axis and pushes fluid out in the perpendicular, see Fig.~\ref{FlowStream}a.i,ii. The swimmer swims small-sphere-leading in this $\text{Re}$ range. For $\text{Re} > 15$, the swimmer switches direction and swims with the large sphere on the front, while the flow field is more pusher-like, i.e. it pushes fluid away along the swimming axis and pulls fluid in on the perpendicular, see Fig.~\ref{FlowStream}a.iii,iv and references~\cite{dombrowski2019transition,dombrowski2020kinematics}. Based on these findings, we might expect there to be an attraction along the swimming axis and a repulsion perpendicular when $\text{Re} < 15$. On the other hand, when $\text{Re} > 15$, we might expect there to be a repulsion along the swimming axis and an attraction on the perpendicular. What we will show next is that while sometimes this is true, the pair system is more complex and interesting. 

We calculated the stroke-averaged fluid flow for the four stable pairs after they reached a steady state, over the range $0.5 \leq \text{Re} \leq 50$. We show four characteristic Reynolds numbers, $\text{Re}$ =  0.5, 5, 20, and 35, that capture most of the interesting behavior, see Fig.~\ref{FlowStream}(b-e). For comparison, we also show the stroke-averaged fluid flow for a single swimmer at the same $\text{Re}$, see Fig.~\ref{FlowStream}(a).The way each pair was initialized is reported in the Supplemental Information (Section III). 

Consider the in-tandem V pair as a function of $\text{Re}$. For $Re>0.5$, once the swimmers found their preferred relative positions, the pair swum together in-tandem as one. As $\text{Re}$ increases, both the angle and distance between the swimmers get smaller, see Fig.~\ref{FlowStream}b.ii-iv and Supplemental Information (Section III, Fig.S3). The flow field around the pair for $\text{Re} = 5$ is qualitatively different from the puller-like flow field around a single swimmer at the same $\text{Re}$, see Figs.~\ref{FlowStream}b.ii and a.ii respectively. 
Specifically, the flow field for the pair at $\text{Re}=5$ pulls in along the swimming axis from underneath the small spheres, pulls through the pair like a zipline, and pushes out behind the large spheres, while elongated vortices return the flow back towards the small spheres. Yet, the pair is similar to the single swimmer in that they both swim small-sphere leading. 
At $\text{Re} = 20$, the flow field of the pair resembles a pusher pushing fluid out along the swimming axis and pulling fluid in towards the perpendicular, though the field is not quite as symmetric as in the single swimmer case (also pusher-like), see Fig.~\ref{FlowStream}b.iii and a.iii or as in the V-pair at lower Re. In fact, we see an arc to the V-pair's trajectory around $\text{Re}\approx 20$, which is the reason why the translational velocity plateaus between $\text{Re}=20-30$, see Fig.3. For $\text{Re} \geq 35$, the pair swims large-sphere-leading (as does the single swimmer) but there is an extra periodicity to the swim stroke. The pair stagger, switch leading swimmers, and shed vortices, Fig.~\ref{FlowStream}a.iv. For a more dynamic visualization, see the movie provided in the Supplemental Information.
Interestingly, for $\text{Re}\leq 0.5$, the swimmers do not really form a V pair, as the angle between them becomes $\theta = 180$. The pair forms a line with the small spheres closest, and remains stationary overall, like a pump, see Fig.~\ref{FlowStream}b.i. The symmetry of the pump pair is reflected in the flow field which pulls in along the line connecting the spheres of each swimmer and pushes away in the perpendicular (puller-like). Note that each swimmer individually would be swimming (albeit slowly) small-sphere-leading, see Fig.~\ref{FlowStream}a.i. Thus, it seems that two swimmers together interacted in such a way that they formed an immotile pair.

\begin{figure*}
\includegraphics[width=1\textwidth]{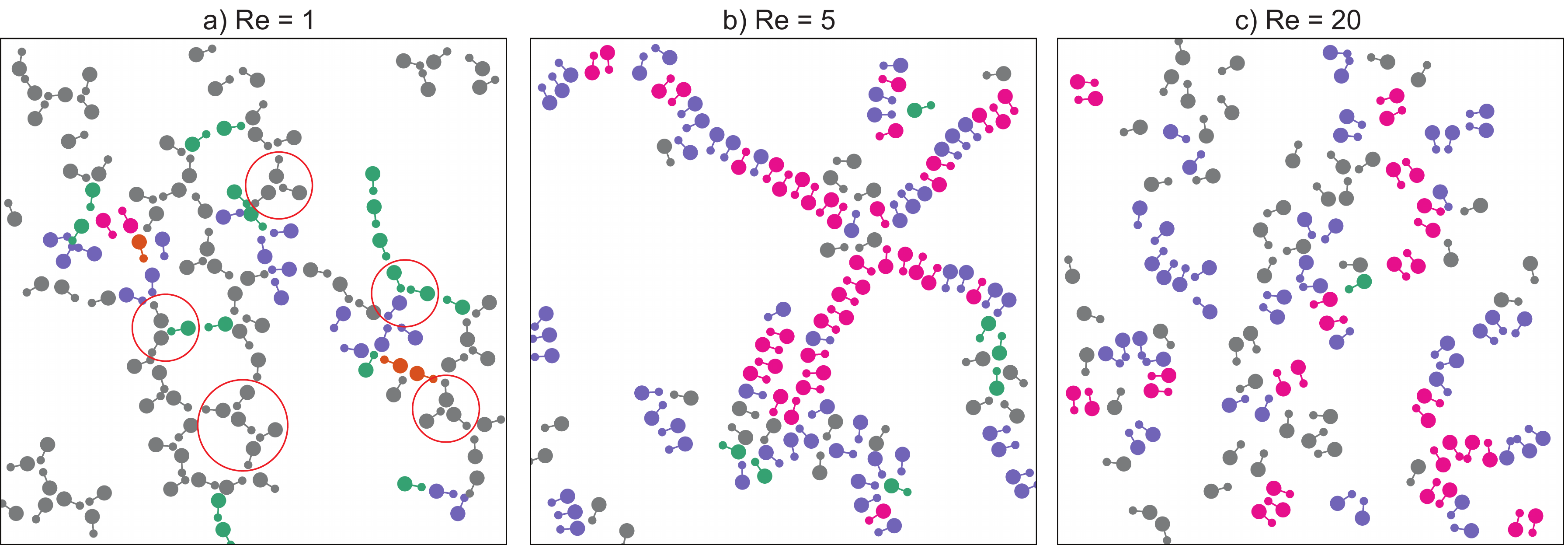}
\centering
\caption{Snapshots of a system composed of 122 swimmers at a) $\text{Re} = 1$, b) $\text{Re} = 5$, and c) $\text{Re} = 20$. 
%three different Reynolds numbers Re = 1, 5, and 20. 
Swimmers are colored according to which stable pair configuration they belong: green for parallel in-line, orange for anti-parallel in-line, purple for in-tandem V, pink for in-tandem orbit, and gray for other. Note that if a swimmer appears in more than one stable pair, then its color is of the first pair identified. a) Swimmers organize in a diffuse network-like structure wherein many swimmers are in in-tandem V or parallel in-line configurations. b) Most swimmers organize themselves in in-tandem orbit or in-tandem V pairs making up the network strands. c) Many swimmers are identified to form in-tandem V or in-tandem orbit pairs; however, they mostly appear in few large clusters, in sharp contrast to the network-like structures seen at $\text{Re} = 1$ and $5$. Open red circles in (a) highlight some triple swimmers with the angles of 120 degree}
\label{PairColl}
\end{figure*}

We next studied the stroke-averaged fluid flows for the in-tandem orbit pair, see Fig.~\ref{FlowStream}c. When $\text{Re} \leq 6$, the swimmers swim away from one another, in a small-sphere-leading spiral, see Fig.~\ref{FlowStream}c.i-ii. When $6.5 \leq \text{Re} \leq 30$, the swimming pair attract enough to form a large-sphere-leading stable orbit about their combined center of mass, as shown in Fig.~\ref{FlowStream}c.iii. Again, we see that having two interacting swimmers instead of a single one changes the dynamics: while the single swimmer transitions from small-sphere-leading to large-sphere-leading at $\text{Re} = 15$, the orbiting pair makes an analogous transition from a small-sphere-leading diverging spiral to a large-sphere-leading stable orbit at a lower $\text{Re} = 6.5$. 
As $\text{Re}$ increases (e.g. $\text{Re} = 20$), when the orbiting pair rotates quickly enough, the outer set of vortices shed, and due to the rotation, the shed vortices seem to recombine with the following outer vortex, see Fig.~\ref{FlowStream}c.iii and movie. 
For $\text{Re} > 30$, the swimming pair physically collide and are unable to maintain a stable orbit, see Fig.~\ref{FlowStream}c.iv.

The parallel in-line pair attract and swim together as one, and the behavior of the pair is similar to that of a single swimmer: both swim small-sphere-leading and generate a flow field that is puller-like (Fig.~\ref{FlowStream}d.i,d.ii and a.i, a.ii). For $Re > 7.5$, the swimmers collide and push away from each other, see Fig.~\ref{FlowStream}d.iii,d.iv.

When $0.5 \leq \text{Re} < 10$, for the antiparallel in-line (head-on) pair, the swimmers aligned along their swimming axis with their large spheres closest and formed a net-motionless pump pair, see Fig.~\ref{FlowStream}e.i-ii. At these Reynolds numbers, each swimmer would individually swim small-sphere-leading. But as they interact with one another in pairs, they get ``trapped'' and form a symmetric pair that stays in place, like a pump, the fluid flows of which resemble a puller. As $\text{Re}$ increases, the swimmers approach one another enough to the point of collision. When $\text{Re} > 15$, the swimmers individually swim large-sphere-leading, which results in the pair periodically colliding into one another as their swimming paths are blocked by each other, see Fig.~\ref{FlowStream}e.iii. Eventually, for $\text{Re} \geq 35$, the pair collide hard enough that they swim away from one another, see Fig.~\ref{FlowStream}e.iv.

\section{Collective Behavior of Multiple Swimmers}
Our main goal for studying pair interactions was to build on our findings and explore whether from pairwise interactions we can explain the collective behavior of multiple swimmers. We note that the collective behavior of intermediate $\text{Re}$ swimmers is largely unknown.

We studied a system of 122 identical swimmers over the Reynolds number range $0.1 \leq \text{Re} \leq 40$, and present three representative cases for $\text{Re} = 1$, $5$, and $20$. We monitored the dumbbell positions over the course of 1000 oscillations and developed an algorithm to identify stable pairs (from the previous section) that form here, too. All systems reached a statistical (and dynamical) steady state after 300 oscillations, where there is no significant change in their topology with time as confirmed by visual inspection. The pair identification algorithm is detailed in the Supplemental Information (Section V). Fig.~\ref{PairColl} shows a snapshot of the collective behavior after 500 oscillations for $\text{Re} = 1, 5, $ and $20$. The swimmers are colored based on the first identified stable pair. Note that swimmers may appear in more than one stable pair. 

When $\text{Re} = 1$, the swimmers organize in a network-like structure with loops and nodes, see Fig.~\ref{PairColl}a. The network is composed of many parallel in-line pair strands (green), as well as in-tandem V pairs (blue). The identified pairs are consistent with the pairs we expect to see at $\text{Re} = 1$. Interestingly, the swimmers also form a structure made up of three swimmers with angles of $\theta \approx 120$. This three-body interaction is found at a lot of the nodes of the network, see red circles in Fig.~\ref{PairColl}a, and of course could not have been predicted by pairwise interactions. 

When $\text{Re} = 5$, a closer-packed network develops consisting of in-tandem V pairs and anti-parallel in-tandem orbits, though because there are many more swimmers they do not really orbit, see Fig.~\ref{PairColl}b (pink and blue). We notice the stable pairs dominate the collective structure, and the in-line pairs are less frequent than at $\text{Re} = 1$. There is thus a clear transition from in-line pairs to in-tandem pairs as Re increases, which is what we saw with the pairs, too. In other words, as the Reynolds number increases, we see a transition in the collective behavior from network-like in-line structures to in-tandem clustering. When $\text{Re} = 20$, the swimmers develop into localized, transient in-tandem clusters which exchange members through physical collisions, see Fig.~\ref{PairColl}c. 

The overall trend from in-line network-like connections to in-tandem clusters as $\text{Re}$ increases seems to hold for both pairs and multiple swimmers. The details, as well as, many-body interactions also indicate more complex behavior for the multiple swimmer case, such as the triples in Fig.~\ref{PairColl}a.

\section{Conclusions}

We computationally studied the pair interactions and collective behavior of asymmetric, dumbbell swimmers over a range of intermediate Reynolds numbers and initial configurations. We found rich and complex pairwise interactions that assembled four stable pairs: in-line and in-tandem, parallel and anti-parallel. We compared the stable pairs' speeds, swim direction and fluid flows to those of the single swimmer. The in-line stable pairs behaved much like the single swimmer transitioning from puller-like to pusher-like stroke-averaged flow fields. In contrast, for the in-tandem pairs we discovered differences in the swim direction transition, as well as the stroke-averaged fluid flow directions. Notably, the in-tandem V pair switched its swim direction at a higher $\text{Re}$ than the single swimmer while the in-tandem orbiting pair switched at a lower $\text{Re}$. We also studied a system of 122 swimmers and found the collective behavior transitioned from in-line network-like connections to small, transient in-tandem clusters as the Reynolds number increased, consistent with the in-line to in-tandem pairwise behavior. Details in the collective behavior involved the formation of triples, and other many-body hydrodynamic interactions that were not predicted or captured by the pair or the single-swimmer behavior.

Our findings show that even a simple model swimmer can demonstrate the richness of intermediate $\text{Re}$ dynamics and collective behavior. 
The fact that the swim direction transitions between small-sphere-leading to large-sphere-leading at different Re’s for the different stable pairs and the single swimmer is an interesting finding that leads to several insights. 
It demonstrates the complexity of pairwise hydrodynamic interactions at intermediate $\text{Re}$ and how behavior changes within the intermediate-$\text{Re}$.
 It indicates that pairwise coordinated swimming may be beneficial hydrodynamically versus isolated swimming. The pairs' behavior as well as the multiple swimmers assemblies suggest that an effective Reynolds number (or other dimensionless ratio) which accounts for the number of swimmers might be a useful tool for describing many-body systems. In the future, it would be interesting to determine specifically possible benefits behind forming stable pairs or multiple-swimmer assemblies versus isolated swimming, e.g. by measuring and comparing efficiencies, and stresses.

\newpage
%\acknowledgements
\textbf{Acknowledgments}
D.K. and T.D acknowledge the National Science Foundation, grant award DMR-1753148.

\bibliography{bibliography}

%apsrev4-2.bst 2018-12-27 (MD) hand-edited version of apsrev4-1.bst
\begin{thebibliography}{27}%
\makeatletter
\providecommand \@ifxundefined [1]{%
 \@ifx{#1\undefined}
}%
\providecommand \@ifnum [1]{%
 \ifnum #1\expandafter \@firstoftwo
 \else \expandafter \@secondoftwo
 \fi
}%
\providecommand \@ifx [1]{%
 \ifx #1\expandafter \@firstoftwo
 \else \expandafter \@secondoftwo
 \fi
}%
\providecommand \natexlab [1]{#1}%
\providecommand \enquote  [1]{``#1''}%
\providecommand \bibnamefont  [1]{#1}%
\providecommand \bibfnamefont [1]{#1}%
\providecommand \citenamefont [1]{#1}%
\providecommand \href@noop [0]{\@secondoftwo}%
\providecommand \href [0]{\begingroup \@sanitize@url \@href}%
\providecommand \@href[1]{\@@startlink{#1}\@@href}%
\providecommand \@@href[1]{\endgroup#1\@@endlink}%
\providecommand \@sanitize@url [0]{\catcode `\\12\catcode `\$12\catcode
  `\&12\catcode `\#12\catcode `\^12\catcode `\_12\catcode `\%12\relax}%
\providecommand \@@startlink[1]{}%
\providecommand \@@endlink[0]{}%
\providecommand \url  [0]{\begingroup\@sanitize@url \@url }%
\providecommand \@url [1]{\endgroup\@href {#1}{\urlprefix }}%
\providecommand \urlprefix  [0]{URL }%
\providecommand \Eprint [0]{\href }%
\providecommand \doibase [0]{https://doi.org/}%
\providecommand \selectlanguage [0]{\@gobble}%
\providecommand \bibinfo  [0]{\@secondoftwo}%
\providecommand \bibfield  [0]{\@secondoftwo}%
\providecommand \translation [1]{[#1]}%
\providecommand \BibitemOpen [0]{}%
\providecommand \bibitemStop [0]{}%
\providecommand \bibitemNoStop [0]{.\EOS\space}%
\providecommand \EOS [0]{\spacefactor3000\relax}%
\providecommand \BibitemShut  [1]{\csname bibitem#1\endcsname}%
\let\auto@bib@innerbib\@empty
%</preamble>
\bibitem [{\citenamefont {Xu}\ \emph {et~al.}(2019)\citenamefont {Xu},
  \citenamefont {Dauparas}, \citenamefont {Das}, \citenamefont {Lauga},\ and\
  \citenamefont {Wu}}]{xu2019self}%
  \BibitemOpen
  \bibfield  {author} {\bibinfo {author} {\bibfnamefont {H.}~\bibnamefont
  {Xu}}, \bibinfo {author} {\bibfnamefont {J.}~\bibnamefont {Dauparas}},
  \bibinfo {author} {\bibfnamefont {D.}~\bibnamefont {Das}}, \bibinfo {author}
  {\bibfnamefont {E.}~\bibnamefont {Lauga}}, and\ \bibinfo {author}
  {\bibfnamefont {Y.}~\bibnamefont {Wu}},\ }\bibfield  {title} {\bibinfo
  {title} {Self-organization of swimmers drives long-range fluid transport in
  bacterial colonies},\ }\href@noop {} {\bibfield  {journal} {\bibinfo
  {journal} {Nature communications}\ }\textbf {\bibinfo {volume} {10}},\
  \bibinfo {pages} {1} (\bibinfo {year} {2019})}\BibitemShut {NoStop}%
\bibitem [{\citenamefont {Houghton}\ \emph {et~al.}(2018)\citenamefont
  {Houghton}, \citenamefont {Koseff}, \citenamefont {Monismith},\ and\
  \citenamefont {Dabiri}}]{houghton2018vertically}%
  \BibitemOpen
  \bibfield  {author} {\bibinfo {author} {\bibfnamefont {I.~A.}\ \bibnamefont
  {Houghton}}, \bibinfo {author} {\bibfnamefont {J.~R.}\ \bibnamefont
  {Koseff}}, \bibinfo {author} {\bibfnamefont {S.~G.}\ \bibnamefont
  {Monismith}}, and\ \bibinfo {author} {\bibfnamefont {J.~O.}\ \bibnamefont
  {Dabiri}},\ }\bibfield  {title} {\bibinfo {title} {Vertically migrating
  swimmers generate aggregation-scale eddies in a stratified column},\
  }\href@noop {} {\bibfield  {journal} {\bibinfo  {journal} {Nature}\ }\textbf
  {\bibinfo {volume} {556}},\ \bibinfo {pages} {497} (\bibinfo {year}
  {2018})}\BibitemShut {NoStop}%
\bibitem [{\citenamefont {Chisholm}\ and\ \citenamefont
  {Khair}(2018)}]{chisholm2018partial}%
  \BibitemOpen
  \bibfield  {author} {\bibinfo {author} {\bibfnamefont {N.~G.}\ \bibnamefont
  {Chisholm}}and\ \bibinfo {author} {\bibfnamefont {A.~S.}\ \bibnamefont
  {Khair}},\ }\bibfield  {title} {\bibinfo {title} {Partial drift volume due to
  a self-propelled swimmer},\ }\href@noop {} {\bibfield  {journal} {\bibinfo
  {journal} {Physical Review Fluids}\ }\textbf {\bibinfo {volume} {3}},\
  \bibinfo {pages} {014501} (\bibinfo {year} {2018})}\BibitemShut {NoStop}%
\bibitem [{\citenamefont {Wilhelmus}\ and\ \citenamefont
  {Dabiri}(2014)}]{wilhelmus2014observations}%
  \BibitemOpen
  \bibfield  {author} {\bibinfo {author} {\bibfnamefont {M.~M.}\ \bibnamefont
  {Wilhelmus}}and\ \bibinfo {author} {\bibfnamefont {J.~O.}\ \bibnamefont
  {Dabiri}},\ }\bibfield  {title} {\bibinfo {title} {Observations of
  large-scale fluid transport by laser-guided plankton aggregations a)},\
  }\href@noop {} {\bibfield  {journal} {\bibinfo  {journal} {Physics of
  Fluids}\ }\textbf {\bibinfo {volume} {26}},\ \bibinfo {pages} {101302}
  (\bibinfo {year} {2014})}\BibitemShut {NoStop}%
\bibitem [{\citenamefont {Kelley}\ and\ \citenamefont
  {Ouellette}(2013)}]{kelley2013emergent}%
  \BibitemOpen
  \bibfield  {author} {\bibinfo {author} {\bibfnamefont {D.~H.}\ \bibnamefont
  {Kelley}}and\ \bibinfo {author} {\bibfnamefont {N.~T.}\ \bibnamefont
  {Ouellette}},\ }\bibfield  {title} {\bibinfo {title} {Emergent dynamics of
  laboratory insect swarms},\ }\href@noop {} {\bibfield  {journal} {\bibinfo
  {journal} {Scientific reports}\ }\textbf {\bibinfo {volume} {3}},\ \bibinfo
  {pages} {1} (\bibinfo {year} {2013})}\BibitemShut {NoStop}%
\bibitem [{\citenamefont {Purcell}(1977)}]{Purcell1977}%
  \BibitemOpen
  \bibfield  {author} {\bibinfo {author} {\bibfnamefont {E.~M.}\ \bibnamefont
  {Purcell}},\ }\bibfield  {title} {\bibinfo {title} {{Life at low Reynolds
  number}},\ }\href {https://doi.org/10.1119/1.10903} {\bibfield  {journal}
  {\bibinfo  {journal} {American Journal of Physics}\ }\textbf {\bibinfo
  {volume} {45}},\ \bibinfo {pages} {3} (\bibinfo {year} {1977})}\BibitemShut
  {NoStop}%
\bibitem [{\citenamefont {Sokolov}\ \emph {et~al.}(2015)\citenamefont
  {Sokolov}, \citenamefont {Zhou}, \citenamefont {Lavrentovich},\ and\
  \citenamefont {Aranson}}]{sokolov2015individual}%
  \BibitemOpen
  \bibfield  {author} {\bibinfo {author} {\bibfnamefont {A.}~\bibnamefont
  {Sokolov}}, \bibinfo {author} {\bibfnamefont {S.}~\bibnamefont {Zhou}},
  \bibinfo {author} {\bibfnamefont {O.~D.}\ \bibnamefont {Lavrentovich}}, and\
  \bibinfo {author} {\bibfnamefont {I.~S.}\ \bibnamefont {Aranson}},\
  }\bibfield  {title} {\bibinfo {title} {Individual behavior and pairwise
  interactions between microswimmers in anisotropic liquid},\ }\href@noop {}
  {\bibfield  {journal} {\bibinfo  {journal} {Physical Review E}\ }\textbf
  {\bibinfo {volume} {91}},\ \bibinfo {pages} {013009} (\bibinfo {year}
  {2015})}\BibitemShut {NoStop}%
\bibitem [{\citenamefont {Elfring}\ and\ \citenamefont
  {Lauga}(2011)}]{elfring2011passive}%
  \BibitemOpen
  \bibfield  {author} {\bibinfo {author} {\bibfnamefont {G.~J.}\ \bibnamefont
  {Elfring}}and\ \bibinfo {author} {\bibfnamefont {E.}~\bibnamefont {Lauga}},\
  }\bibfield  {title} {\bibinfo {title} {Passive hydrodynamic synchronization
  of two-dimensional swimming cells},\ }\href@noop {} {\bibfield  {journal}
  {\bibinfo  {journal} {Physics of Fluids}\ }\textbf {\bibinfo {volume} {23}},\
  \bibinfo {pages} {011902} (\bibinfo {year} {2011})}\BibitemShut {NoStop}%
\bibitem [{\citenamefont {Yang}\ \emph {et~al.}(2008)\citenamefont {Yang},
  \citenamefont {Elgeti},\ and\ \citenamefont {Gompper}}]{yang2008cooperation}%
  \BibitemOpen
  \bibfield  {author} {\bibinfo {author} {\bibfnamefont {Y.}~\bibnamefont
  {Yang}}, \bibinfo {author} {\bibfnamefont {J.}~\bibnamefont {Elgeti}}, and\
  \bibinfo {author} {\bibfnamefont {G.}~\bibnamefont {Gompper}},\ }\bibfield
  {title} {\bibinfo {title} {Cooperation of sperm in two dimensions:
  synchronization, attraction, and aggregation through hydrodynamic
  interactions},\ }\href@noop {} {\bibfield  {journal} {\bibinfo  {journal}
  {Physical review E}\ }\textbf {\bibinfo {volume} {78}},\ \bibinfo {pages}
  {061903} (\bibinfo {year} {2008})}\BibitemShut {NoStop}%
\bibitem [{\citenamefont {Ishikawa}\ \emph {et~al.}(2006)\citenamefont
  {Ishikawa}, \citenamefont {Simmonds},\ and\ \citenamefont
  {Pedley}}]{ishikawa2006hydrodynamic}%
  \BibitemOpen
  \bibfield  {author} {\bibinfo {author} {\bibfnamefont {T.}~\bibnamefont
  {Ishikawa}}, \bibinfo {author} {\bibfnamefont {M.}~\bibnamefont {Simmonds}},
  and\ \bibinfo {author} {\bibfnamefont {T.~J.}\ \bibnamefont {Pedley}},\
  }\bibfield  {title} {\bibinfo {title} {Hydrodynamic interaction of two
  swimming model micro-organisms},\ }\href@noop {} {\bibfield  {journal}
  {\bibinfo  {journal} {Journal of Fluid Mechanics}\ }\textbf {\bibinfo
  {volume} {568}},\ \bibinfo {pages} {119} (\bibinfo {year}
  {2006})}\BibitemShut {NoStop}%
\bibitem [{\citenamefont {Kanevsky}\ \emph {et~al.}(2010)\citenamefont
  {Kanevsky}, \citenamefont {Shelley},\ and\ \citenamefont
  {Tornberg}}]{kanevsky2010modeling}%
  \BibitemOpen
  \bibfield  {author} {\bibinfo {author} {\bibfnamefont {A.}~\bibnamefont
  {Kanevsky}}, \bibinfo {author} {\bibfnamefont {M.~J.}\ \bibnamefont
  {Shelley}}, and\ \bibinfo {author} {\bibfnamefont {A.-K.}\ \bibnamefont
  {Tornberg}},\ }\bibfield  {title} {\bibinfo {title} {Modeling simple
  locomotors in stokes flow},\ }\href@noop {} {\bibfield  {journal} {\bibinfo
  {journal} {Journal of Computational Physics}\ }\textbf {\bibinfo {volume}
  {229}},\ \bibinfo {pages} {958} (\bibinfo {year} {2010})}\BibitemShut
  {NoStop}%
\bibitem [{\citenamefont {Lippera}\ \emph {et~al.}(2020)\citenamefont
  {Lippera}, \citenamefont {Benzaquen},\ and\ \citenamefont
  {Michelin}}]{lippera2020bouncing}%
  \BibitemOpen
  \bibfield  {author} {\bibinfo {author} {\bibfnamefont {K.}~\bibnamefont
  {Lippera}}, \bibinfo {author} {\bibfnamefont {M.}~\bibnamefont {Benzaquen}},
  and\ \bibinfo {author} {\bibfnamefont {S.}~\bibnamefont {Michelin}},\
  }\bibfield  {title} {\bibinfo {title} {Bouncing, chasing, or pausing:
  Asymmetric collisions of active droplets},\ }\href@noop {} {\bibfield
  {journal} {\bibinfo  {journal} {Physical Review Fluids}\ }\textbf {\bibinfo
  {volume} {5}},\ \bibinfo {pages} {032201} (\bibinfo {year}
  {2020})}\BibitemShut {NoStop}%
\bibitem [{\citenamefont {Lauga}\ and\ \citenamefont
  {Powers}(2009)}]{lauga2009hydrodynamics}%
  \BibitemOpen
  \bibfield  {author} {\bibinfo {author} {\bibfnamefont {E.}~\bibnamefont
  {Lauga}}and\ \bibinfo {author} {\bibfnamefont {T.~R.}\ \bibnamefont
  {Powers}},\ }\bibfield  {title} {\bibinfo {title} {The hydrodynamics of
  swimming microorganisms},\ }\href@noop {} {\bibfield  {journal} {\bibinfo
  {journal} {Reports on Progress in Physics}\ }\textbf {\bibinfo {volume}
  {72}},\ \bibinfo {pages} {096601} (\bibinfo {year} {2009})}\BibitemShut
  {NoStop}%
\bibitem [{\citenamefont {Lippera}\ \emph {et~al.}(2021)\citenamefont
  {Lippera}, \citenamefont {Benzaquen},\ and\ \citenamefont
  {Michelin}}]{lippera2021alignment}%
  \BibitemOpen
  \bibfield  {author} {\bibinfo {author} {\bibfnamefont {K.}~\bibnamefont
  {Lippera}}, \bibinfo {author} {\bibfnamefont {M.}~\bibnamefont {Benzaquen}},
  and\ \bibinfo {author} {\bibfnamefont {S.}~\bibnamefont {Michelin}},\
  }\bibfield  {title} {\bibinfo {title} {Alignment and scattering of colliding
  active droplets},\ }\href@noop {} {\bibfield  {journal} {\bibinfo  {journal}
  {Soft Matter}\ }\textbf {\bibinfo {volume} {17}},\ \bibinfo {pages} {365}
  (\bibinfo {year} {2021})}\BibitemShut {NoStop}%
\bibitem [{\citenamefont {Koiller}\ \emph {et~al.}(1996)\citenamefont
  {Koiller}, \citenamefont {Ehlers},\ and\ \citenamefont
  {Montgomery}}]{koiller1996problems}%
  \BibitemOpen
  \bibfield  {author} {\bibinfo {author} {\bibfnamefont {J.}~\bibnamefont
  {Koiller}}, \bibinfo {author} {\bibfnamefont {K.}~\bibnamefont {Ehlers}},
  and\ \bibinfo {author} {\bibfnamefont {R.}~\bibnamefont {Montgomery}},\
  }\bibfield  {title} {\bibinfo {title} {Problems and progress in
  microswimming},\ }\href@noop {} {\bibfield  {journal} {\bibinfo  {journal}
  {Journal of Nonlinear Science}\ }\textbf {\bibinfo {volume} {6}},\ \bibinfo
  {pages} {507} (\bibinfo {year} {1996})}\BibitemShut {NoStop}%
\bibitem [{\citenamefont {Alexander}\ and\ \citenamefont
  {Yeomans}(2008)}]{alexander2008dumb}%
  \BibitemOpen
  \bibfield  {author} {\bibinfo {author} {\bibfnamefont {G.}~\bibnamefont
  {Alexander}}and\ \bibinfo {author} {\bibfnamefont {J.}~\bibnamefont
  {Yeomans}},\ }\bibfield  {title} {\bibinfo {title} {Dumb-bell swimmers},\
  }\href@noop {} {\bibfield  {journal} {\bibinfo  {journal} {EPL (Europhysics
  Letters)}\ }\textbf {\bibinfo {volume} {83}},\ \bibinfo {pages} {34006}
  (\bibinfo {year} {2008})}\BibitemShut {NoStop}%
\bibitem [{\citenamefont {Lauga}\ and\ \citenamefont
  {Bartolo}(2008)}]{lauga2008no}%
  \BibitemOpen
  \bibfield  {author} {\bibinfo {author} {\bibfnamefont {E.}~\bibnamefont
  {Lauga}}and\ \bibinfo {author} {\bibfnamefont {D.}~\bibnamefont {Bartolo}},\
  }\bibfield  {title} {\bibinfo {title} {No many-scallop theorem: Collective
  locomotion of reciprocal swimmers},\ }\href@noop {} {\bibfield  {journal}
  {\bibinfo  {journal} {Physical Review E}\ }\textbf {\bibinfo {volume} {78}},\
  \bibinfo {pages} {030901} (\bibinfo {year} {2008})}\BibitemShut {NoStop}%
\bibitem [{\citenamefont {Putz}\ and\ \citenamefont {Dunkel}(2010)}]{Putz2010}%
  \BibitemOpen
  \bibfield  {author} {\bibinfo {author} {\bibfnamefont {V.~B.}\ \bibnamefont
  {Putz}}and\ \bibinfo {author} {\bibfnamefont {J.}~\bibnamefont {Dunkel}},\
  }\bibfield  {title} {\bibinfo {title} {{Low Reynolds number hydrodynamics of
  asymmetric, oscillating dumbbell pairs}},\ }\href
  {https://doi.org/10.1140/epjst/e2010-01278-y} {\bibfield  {journal} {\bibinfo
   {journal} {The European Physical Journal Special Topics}\ }\textbf {\bibinfo
  {volume} {187}},\ \bibinfo {pages} {135} (\bibinfo {year}
  {2010})}\BibitemShut {NoStop}%
\bibitem [{\citenamefont {Dombrowski}\ \emph {et~al.}(2019)\citenamefont
  {Dombrowski}, \citenamefont {Jones}, \citenamefont {Katsikis}, \citenamefont
  {Bhalla}, \citenamefont {Griffith},\ and\ \citenamefont
  {Klotsa}}]{dombrowski2019transition}%
  \BibitemOpen
  \bibfield  {author} {\bibinfo {author} {\bibfnamefont {T.}~\bibnamefont
  {Dombrowski}}, \bibinfo {author} {\bibfnamefont {S.~K.}\ \bibnamefont
  {Jones}}, \bibinfo {author} {\bibfnamefont {G.}~\bibnamefont {Katsikis}},
  \bibinfo {author} {\bibfnamefont {A.~P.~S.}\ \bibnamefont {Bhalla}}, \bibinfo
  {author} {\bibfnamefont {B.~E.}\ \bibnamefont {Griffith}}, and\ \bibinfo
  {author} {\bibfnamefont {D.}~\bibnamefont {Klotsa}},\ }\bibfield  {title}
  {\bibinfo {title} {Transition in swimming direction in a model self-propelled
  inertial swimmer},\ }\href@noop {} {\bibfield  {journal} {\bibinfo  {journal}
  {Physical Review Fluids}\ }\textbf {\bibinfo {volume} {4}},\ \bibinfo {pages}
  {021101} (\bibinfo {year} {2019})}\BibitemShut {NoStop}%
\bibitem [{\citenamefont {Dombrowski}\ and\ \citenamefont
  {Klotsa}(2020)}]{dombrowski2020kinematics}%
  \BibitemOpen
  \bibfield  {author} {\bibinfo {author} {\bibfnamefont {T.}~\bibnamefont
  {Dombrowski}}and\ \bibinfo {author} {\bibfnamefont {D.}~\bibnamefont
  {Klotsa}},\ }\bibfield  {title} {\bibinfo {title} {Kinematics of a simple
  reciprocal model swimmer at intermediate reynolds numbers},\ }\href@noop {}
  {\bibfield  {journal} {\bibinfo  {journal} {Physical Review Fluids}\ }\textbf
  {\bibinfo {volume} {5}},\ \bibinfo {pages} {063103} (\bibinfo {year}
  {2020})}\BibitemShut {NoStop}%
\bibitem [{\citenamefont {Gemmell}\ \emph {et~al.}(2013)\citenamefont
  {Gemmell}, \citenamefont {Sheng},\ and\ \citenamefont
  {Buskey}}]{gemmell2013compensatory}%
  \BibitemOpen
  \bibfield  {author} {\bibinfo {author} {\bibfnamefont {B.~J.}\ \bibnamefont
  {Gemmell}}, \bibinfo {author} {\bibfnamefont {J.}~\bibnamefont {Sheng}}, and\
  \bibinfo {author} {\bibfnamefont {E.~J.}\ \bibnamefont {Buskey}},\ }\bibfield
   {title} {\bibinfo {title} {Compensatory escape mechanism at low reynolds
  number},\ }\href@noop {} {\bibfield  {journal} {\bibinfo  {journal}
  {Proceedings of the National Academy of Sciences}\ }\textbf {\bibinfo
  {volume} {110}},\ \bibinfo {pages} {4661} (\bibinfo {year}
  {2013})}\BibitemShut {NoStop}%
\bibitem [{\citenamefont {G{\"o}tze}\ and\ \citenamefont
  {Gompper}(2010)}]{gotze2010mesoscale}%
  \BibitemOpen
  \bibfield  {author} {\bibinfo {author} {\bibfnamefont {I.~O.}\ \bibnamefont
  {G{\"o}tze}}and\ \bibinfo {author} {\bibfnamefont {G.}~\bibnamefont
  {Gompper}},\ }\bibfield  {title} {\bibinfo {title} {Mesoscale simulations of
  hydrodynamic squirmer interactions},\ }\href@noop {} {\bibfield  {journal}
  {\bibinfo  {journal} {Physical Review E}\ }\textbf {\bibinfo {volume} {82}},\
  \bibinfo {pages} {041921} (\bibinfo {year} {2010})}\BibitemShut {NoStop}%
\bibitem [{\citenamefont {Li}\ \emph {et~al.}(2016)\citenamefont {Li},
  \citenamefont {Ostace},\ and\ \citenamefont
  {Ardekani}}]{li_arezoodaki_2016hydrodynamic}%
  \BibitemOpen
  \bibfield  {author} {\bibinfo {author} {\bibfnamefont {G.}~\bibnamefont
  {Li}}, \bibinfo {author} {\bibfnamefont {A.}~\bibnamefont {Ostace}}, and\
  \bibinfo {author} {\bibfnamefont {A.~M.}\ \bibnamefont {Ardekani}},\
  }\bibfield  {title} {\bibinfo {title} {Hydrodynamic interaction of swimming
  organisms in an inertial regime},\ }\href@noop {} {\bibfield  {journal}
  {\bibinfo  {journal} {Physical Review E}\ }\textbf {\bibinfo {volume} {94}},\
  \bibinfo {pages} {053104} (\bibinfo {year} {2016})}\BibitemShut {NoStop}%
\bibitem [{\citenamefont {Chatterjee}\ \emph {et~al.}(2019)\citenamefont
  {Chatterjee}, \citenamefont {Rana}, \citenamefont {Simha}, \citenamefont
  {Perlekar},\ and\ \citenamefont {Ramaswamy}}]{chatterjee2019fluid}%
  \BibitemOpen
  \bibfield  {author} {\bibinfo {author} {\bibfnamefont {R.}~\bibnamefont
  {Chatterjee}}, \bibinfo {author} {\bibfnamefont {N.}~\bibnamefont {Rana}},
  \bibinfo {author} {\bibfnamefont {R.~A.}\ \bibnamefont {Simha}}, \bibinfo
  {author} {\bibfnamefont {P.}~\bibnamefont {Perlekar}}, and\ \bibinfo {author}
  {\bibfnamefont {S.}~\bibnamefont {Ramaswamy}},\ }\bibfield  {title} {\bibinfo
  {title} {Fluid flocks with inertia},\ }\href@noop {} {\bibfield  {journal}
  {\bibinfo  {journal} {arXiv preprint arXiv:1907.03492}\ } (\bibinfo {year}
  {2019})}\BibitemShut {NoStop}%
\bibitem [{\citenamefont {Griffith}\ \emph {et~al.}(2007)\citenamefont
  {Griffith}, \citenamefont {Hornung}, \citenamefont {McQueen},\ and\
  \citenamefont {Peskin}}]{griffith2007adaptive}%
  \BibitemOpen
  \bibfield  {author} {\bibinfo {author} {\bibfnamefont {B.~E.}\ \bibnamefont
  {Griffith}}, \bibinfo {author} {\bibfnamefont {R.~D.}\ \bibnamefont
  {Hornung}}, \bibinfo {author} {\bibfnamefont {D.~M.}\ \bibnamefont
  {McQueen}}, and\ \bibinfo {author} {\bibfnamefont {C.~S.}\ \bibnamefont
  {Peskin}},\ }\bibfield  {title} {\bibinfo {title} {{An adaptive, formally
  second order accurate version of the immersed boundary method}},\ }\href@noop
  {} {\bibfield  {journal} {\bibinfo  {journal} {Journal of Computational
  Physics}\ }\textbf {\bibinfo {volume} {223}},\ \bibinfo {pages} {10}
  (\bibinfo {year} {2007})}\BibitemShut {NoStop}%
\bibitem [{\citenamefont {Griffith}(2012)}]{griffith2012immersed}%
  \BibitemOpen
  \bibfield  {author} {\bibinfo {author} {\bibfnamefont {B.~E.}\ \bibnamefont
  {Griffith}},\ }\bibfield  {title} {\bibinfo {title} {Immersed boundary model
  of aortic heart valve dynamics with physiological driving and loading
  conditions},\ }\href@noop {} {\bibfield  {journal} {\bibinfo  {journal}
  {International Journal for Numerical Methods in Biomedical Engineering}\
  }\textbf {\bibinfo {volume} {28}},\ \bibinfo {pages} {317} (\bibinfo {year}
  {2012})}\BibitemShut {NoStop}%
\bibitem [{\citenamefont {Griffith}\ \emph {et~al.}(2009)\citenamefont
  {Griffith}, \citenamefont {Luo}, \citenamefont {McQueen},\ and\ \citenamefont
  {Peskin}}]{Griffith2009Immersed}%
  \BibitemOpen
  \bibfield  {author} {\bibinfo {author} {\bibfnamefont {B.~E.}\ \bibnamefont
  {Griffith}}, \bibinfo {author} {\bibfnamefont {X.}~\bibnamefont {Luo}},
  \bibinfo {author} {\bibfnamefont {D.~M.}\ \bibnamefont {McQueen}}, and\
  \bibinfo {author} {\bibfnamefont {C.~S.}\ \bibnamefont {Peskin}},\ }\bibfield
   {title} {\bibinfo {title} {{Simulating the fluid dynamics of natural and
  prosthetic heart valves using the immersed boundary method}},\ }\href
  {https://doi.org/10.1142/S1758825109000113} {\bibfield  {journal} {\bibinfo
  {journal} {Int. J. Appl. Mech.}\ }\textbf {\bibinfo {volume} {1}},\ \bibinfo
  {pages} {137} (\bibinfo {year} {2009})}\BibitemShut {NoStop}%
\end{thebibliography}%

\end{document}